\documentclass[twocolumn,a4paper,superscriptaddress,prl]{revtex4}
\usepackage{natbib}
\usepackage[dvips]{graphicx}
\begin{document}

\title{The order-parameter symmetry and Fermi surface topology
of 122 Fe-based superconductors: a point-contact Andreev-reflection
study}

\author{R.S. Gonnelli}\author{M. Tortello}\author{D. Daghero}\author{P. Pecchio}\author{S. Galasso}

\affiliation{Dipartimento di Scienza Applicata e Tecnologia, Politecnico di Torino, 10129, Torino, Italy}

\author{V.A. Stepanov}

\affiliation{P. N. Lebedev Physical Institute, Russian Academy of Sciences, 119991, Moscow, Russia}

\author{Z. Bukowski}

\affiliation{Inst. of Low Temp. and Struct. Research, Polish Acad. of Sciences, 50-422, Wroclaw, Poland}

\author{N. D. Zhigadlo}

\affiliation{Laboratory for Solid State Physics, ETHZ, CH-8093, Zurich, Switzerland}

\author{J. Karpinski}

\affiliation{Ecole Polytechnique F\'{e}d\'{e}rale de Lausanne, CH-1015 Lausanne, Switzerland}

\author{K. Iida}\author{B. Holzapfel}

\affiliation{Leibniz Institute for Solid State and Materials Research, 01069 Dresden, Germany}



\begin{abstract}
We report on the results of directional point-contact
Andreev-reflection (PCAR) measurements in
$\mathrm{Ba(Fe_{1-x}Co_x)_2As_2}$ single crystals and epitaxial
$c$-axis oriented films with $x=0.08$ as well as in
$\mathrm{Ca(Fe_{1-x}Co_x)_2As_2}$ single crystals with $x = 0.06$.
The PCAR spectra are analyzed within the two-band 3D version of the
Blonder-Tinkham-Klapwijk model for Andreev reflection we recently
developed, and that makes use of an analytical expression for the
Fermi surface that mimics the one calculated within the
density-functional theory (DFT). The spectra in
$\mathrm{Ca(Fe_{0.94}Co_{0.06})_2As_2}$ unambiguously demonstrate
the presence of nodes or zeros in the small gap. In
$\mathrm{Ba(Fe_{0.92}Co_{0.08})_2As_2}$, the $ab$-plane spectra in
single crystals can be fitted by assuming two nodeless gaps, but
this model fails to fit the $c$-axis ones in epitaxial films. All
these results are discussed in comparison with recent theoretical
predictions about the occurrence of accidental 3D nodes and the
presence of ``hot spots'' in the gaps of 122 compounds.
\end{abstract}


\maketitle

\section{Introduction}
Four years after the discovery of superconductivity in Fe-based
compounds \cite{kamihara08}, a surprising variety of behaviors has
been unveiled not only between compounds of different classes (the
so-called 1111, 11 or 122 classes, just to mention the most studied)
but also among compounds of the same class that share the same
crystal structure. One of the properties that shows unexpected
variations in a given class of Fe-based pnictides is the symmetry of
the order parameter(s) (OPs) so that it is not possible to make a
general statement about the gap symmetry of these compounds. The
multiband electronic structure and the Fermi surface (FS) nesting
that is responsible for the antiferromagnetic instability in the
parent compounds strongly suggest a superconducting pairing mediated
by spin fluctuations \cite{mazin08,mazin10b} which is also supported
by many experimental facts \cite{paglione10}. Various possible
gap symmetries are allowed, at least theoretically, within the
picture of spin-fluctuation mediated pairing. In most of
optimally-doped compounds the so-called $s \pm$ symmetry with
isotropic OPs of opposite sign on the holelike and electronlike FS
sheets is realized \cite{mazin08}, but the emergence of nodes or
zeros on some FSs is theoretically predicted in various situations
\cite{mazin10,kuroki09,suzuki11}.

Here we report on recent PCAR spectroscopy results in single
crystals of 8\% Co-doped BaFe$_2$As$_2$ (Ba-122) with bulk $T_c^{on}=24.5$ K and in
single crystals of 6 \% Co-doped CaFe$_2$As$_2$ (Ca-122) with $T_c^{on}=20.0$ K. In
the first case, the PCAR spectra (taken with the current injected
along the $ab$ planes) are compatible with two nodeless gaps of
different amplitude, the smaller being probably associated to one of
the electrolike FS sheets, in agreement with angle-resolved photoemission spectroscopy (ARPES)
\cite{terashima09}.
In the crystals of Co-doped Ca-122, the PCAR spectra systematically show a zero-bias
peak or maximum, which strongly suggests that the small gap
(residing here on the holelike FS sheet centered at the $\Gamma$
point of the Brillouin zone) presents lines of nodes or zeros on the
FS, while the large gap remains presumably isotropic. This finding
agrees with the predictions about the emergence of 3D nodes in the
OP of 122 compounds when the holelike FS evolves towards a topological
transition from a warped cylinder to separate pockets
\cite{suzuki11,gonnelli12}. Finally, we present new results of PCAR
experiments in epitaxial thin films of 8\% Co-doped Ba-122 (with
$T_c^{mid}=24.35$ K). In this case, the current is injected along
the $c$ axis and the 3D-BTK model with two nodeless gaps fails to
fit the PCAR spectra at low energy. This might be a hallmark of the
presence of ``hot spots'' (where the gap is strongly suppressed) or
3D nodal lines on some of the FSs, which has been proposed
\cite{mazin10} to explain Raman results in this compound
\cite{muschler09}.

\section{Experimental details}
\subsection{The samples}
The $\mathrm{Ba(Fe_{1-x}Co_{x})_2As_2}$ single crystals (with
$x=0.08$) were prepared by the self-flux method \cite{sefat08b}
under a pressure of 280 MPa at the National High Magnetic Field
Laboratory in Tallahassee. The crystal size is $\approx
1\times1\times0.1$ mm$^3$, and the $c$ axis is perpendicular to the
larger surface. The resistive transition sets in at $T_c ^{on}=24.5$
K with $\Delta T_ c$ (10\%-90\%) = 1 K.

The $\mathrm{Ba(Fe_{1-x}Co_{x})_2As_2}$ films (with $x=0.08$) were
deposited at the Leibniz Institute for Solid State and Materials
Research (IFW) in Dresden, Germany. Two kinds of substrates were
used: single-crystalline CaF$_2$ or MgO (in this case, with a Fe
buffer layer on top of it). The films are epitaxial with the $c$
axis normal to the surface.
The superconducting transition of the films on CaF$_2$ has a
midpoint at $T_c^{mid}=24.35$ K, while its width is $\Delta
T_c=T_c^{90\%}-T_c^{10\%}=1.7$ K. For the films on MgO the same
quantities are $T_c^{mid}=23.80$ K and $\Delta T_c=1.5$ K.

The  Ca(Fe$_{1-x}$Co$_x$)$_2$As$_2$ single crystals were grown at
ETH Zurich from Sn flux, as described in Ref. \cite{matusiak10}, and
were plate-like, with the $c$ axis perpendicular to the plate. The
superconducting transition as measured from DC susceptibility sets
in at $T_c^{on}=20$ K and an effective $T_c^{eff}=17$ K can be
determined by extrapolating the linear part of the curve. The
transition is rather broad, but this is common to all the
state-of-the-art crystals of the same compound grown at present
\cite{harnagea11}.
The broad transition is however not detrimental to point-contact
measurements: Simply, the local critical temperature of point
contacts can vary from point to point within the transition width.

\subsection{Point-contact setup and procedure}
The point contacts were made by using the ``soft'' technique
described elsewhere \cite{daghero10}, in which the metallic tip used as
counterelectrode in the more conventional needle-anvil configuration
is replaced by a small drop of Ag paste ($\phi \simeq 50 \,\mu$m) put
on the sample surface. In single crystals the contacts were put on a
fresh side surface (exposed by breaking the crystals) so that the
current was mainly injected along the $ab$ planes. In films, the
point contacts were put on the top surface. Owing to the film
orientation, this means that the probe current was mainly injected
along the $c$ axis.

The conductance curves $\mathrm{d}I/\mathrm{d}V$ vs. $V$ of each
point contact were obtained by numerically differentiating the
measured $I-V$ characteristics. The spectra were then normalized
(i.e. divided by the normal-state conductance of the same contact)
to allow a comparison to the relevant theoretical models. In single
crystals (where the high-bias tails of the spectra at different
temperatures are superimposed to one another) we divided by the conductance curve
recorded just above the critical temperature of the contact, $T_c^A$
(defined as the temperature at which the Andreev reflection features
disappear). In the case of films, this procedure is impossible
because of an abrupt vertical shift of the spectra when the
temperature approaches $T_c^A$. This shift is related to the
transition to the normal state of the
portion of the film between the point contact and the voltage
electrode \cite{chen10b} and has nothing to do with the contact
itself (a detailed model of this phenomenon will be given in a
forthcoming paper). Therefore, the low-temperature experimental
spectra were divided by a guess normal state obtained by fitting
their high-energy tails with a fourth-order polynomial function, as
we did in Ref.\cite{gonnelli09a}.

\section{Calculation of the Fermi surface}

The FS of the materials under study
($\mathrm{Ba(Fe_{1-x}Co_x)_2As_2}$ with $x=0.08$ and
$\mathrm{Ca(Fe_{1-x}Co_x)_2As_2}$ with $x = 0.06$) was calculated
within the Density-Functional Theory (DFT) by using the Elk FP-LAPW
Code ($\mathrm{http://elk.sourceforge.net/}$) and the GGA approach
for the exchange correlation potential. To account for the partial
substitution of Fe with Co we used a virtual-crystal approximation
that indeed works particularly well for the Fe-Co substitution.

\begin{figure}[b]
\includegraphics[width=\columnwidth]{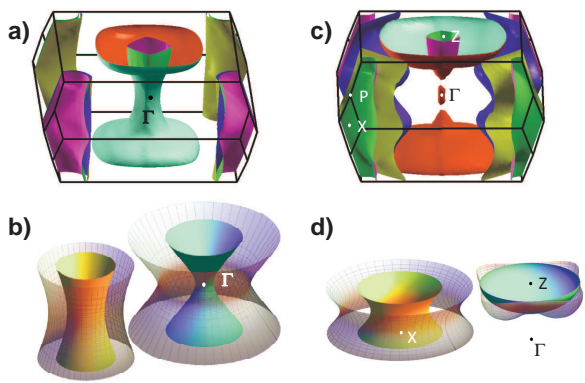}
\caption{(a) The Fermi surface (FS) of $\mathrm{Ba(Fe_{1-x}Co_x)_2As_2}$ with
$x=0.08$. (b) The model FS used in the 3D-BTK model to fit the
experimental conductance curves. Solid surfaces represent the FS
sheets, while gridded surfaces represent the amplitude of the
corresponding energy gap. (c) The FS of
$\mathrm{Ca(Fe_{1-x}Co_x)_2As_2}$ with $x=0.06$. (d) The model FS
(in the upper half of the first Brillouin zone) used in the 3D-BTK model
to fit the experimental conductance curves. Gridded surfaces
represent the amplitudes of the gaps, a large isotropic one on the
electron FS and a small anisotropic one on the hole FS.
}\label{fig:FS}
\end{figure}

The lattice constants used for the calculations in the case of
$\mathrm{Ba(Fe_{1-x}Co_x)_2As_2}$ are $a=b=3.9625$ {\AA} and
$c=13.0168$ {\AA} as in Ref.\cite{singh08b}. The height of the As
atom above the Fe layer is $h_{As}=1.1975$ {\AA}. The resulting FS
is shown in Fig. \ref{fig:FS}(a). It features two holelike FS sheets
at the center of the Brillouin zone (BZ). Both have the shape of
warped cylinders whose cross section is maximum at the top and
bottom edges of the BZ, but the outer one shows a more marked
warping. The situation is somewhat similar to that found in
$\mathrm{BaFe_2As_2}$ with partial substitution of As with P, as
described by Suzuki et al. \cite{suzuki11}. The two electronlike FS
sheets at the corners of the BZ are also warped cylinders with the
characteristic elliptical cross section whose semimajor axis varies
along $k_z$.

In the case of $\mathrm{Ca(Fe_{1-x}Co_x)_2As_2}$, there is no direct
experimental information on the low-temperature lattice constants.
Owing to the small dependence of the room-temperature lattice
parameters on the doping content \cite{hu11}, we assumed the
low-temperature lattice constants of the parent compound
$\mathrm{CaFe_2A_2}$ in the tetragonal phase to be a good first
approximation to the real ones at the doping content of our interest
($x=0.06$). We then started from the lattice constants of the
orthorhombic phase of $\mathrm{CaFe_2A_2}$ at pressure P = 0
calculated as in Ref. \cite{colonna11}, and then we made the
structure tetragonal by averaging $a$ and $b$. The result is
$a=b=3.925$ {\AA} and $c=11.356$ {\AA}. These values are in good
agreement with the experimental ones measured in the tetragonal
phase of $\mathrm{CaFe_2A_2}$ at 300 K (and pressure P = 0.8 - 1 GPa
\cite{mittal11}). Starting from the calculated equilibrium phase and
always considering the antiferromagnetic phase, an optimized
parameter $h_{As}=1.309$ {\AA} was obtained. The charge density was
thus integrated over $8 \times 8\times 4$ $k$ points in the
Brillouin zone and the band structure as well as the FSs were
calculated in the non-magnetic body-centered tetragonal phase. The
resulting FS is shown in Fig. \ref{fig:FS}(c). It is clear that at
this doping content the holelike FS sheets are undergoing a
topological transition. While at lower doping they have the shape of
strongly warped cylinders (similar to those shown for Ba-122 in Fig.
\ref{fig:FS}(a)), at $x=0.06$ they split into separate cup-shaped pockets
centered around the Z points.

\section{Results and analysis}

\subsection{Ba(Fe$_{1-x}$Co$_x$)$_2$As$_2$ single crystals}
\begin{figure}[b]
\includegraphics[width=0.8\columnwidth]{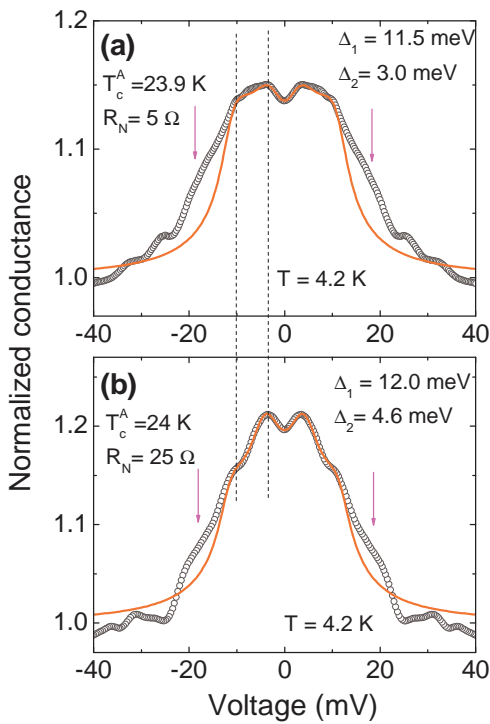}
\caption{Two examples of normalized conductance curves of $ab$-plane
point contacts in $\mathrm{Ba(Fe_{1-x}Co_x)_2As_2}$ single crystals
with $x=0.08$ (symbols) and of the relevant fit (lines) with the 3D BTK
model using the FS shown in Fig. \ref{fig:FS}(b). The fitting
parameters of the curve in (a) are: $\Delta_1=11.5$ meV,
$\Gamma_1=1.85$ meV, $Z_1=0.03$ and $\Delta_2=3.0$ meV,
$\Gamma_2=3.6$ meV, $Z_2=0.310$. The weights of the two bands in the
conductance are $w_1=0.23$ and $w_2=0.77$. The parameters for
the fit of the curve in (b) are instead $\Delta_1=12.0$ meV,
$\Gamma_1=1.75$ meV, $Z_1=0.08$ and $\Delta_2=4.6$ meV,
$\Gamma_2=3.00$ meV, $Z_2=0.245$. The weights in this case are $w_1=
0.21$ and $w_2=0.79$.}\label{fig:Bafit}
\end{figure}

Figure \ref{fig:Bafit} presents two examples of experimental
normalized conductance curves (symbols) of $ab$-plane contacts in
$\mathrm{Ba(Fe_{1-x}Co_x)_2As_2}$ single crystals. Both the curves
in (a) and (b) show a zero-bias dip, two symmetric maxima related to
a small gap, and two other kinks related to a second, larger gap.
Moreover, the spectra feature additional, higher-energy structures
(indicated by arrows) that are likely to be due to the strong
electron-boson coupling \cite{tortello10}. We will not deal with
these structures here; they cannot be reproduced by any BTK model
unless an energy-dependent gap (as obtained from the solution of
Eliashberg equations) is used instead of the constant BCS-like gap
used here \cite{tortello10,daghero11}. To fit the experimental
curves in Fig. \ref{fig:Bafit} we used the 3D version of the
generalized BTK model \cite{daghero11}, that accounts for the real
shape of the FS and is not based on the simplifying assumption that
the latter is spherical or cylindrical, as instead the 1D and 2D BTK
models do. For ease of calculation, we used the model FS shown in
Fig. \ref{fig:FS}(b), which mimics the real one from DFT
calculations reported in Fig. \ref{fig:FS}(a). The former consists of two separate hyperboloids of
revolution, meant to simulate the main holelike and electronlike
sheets, whose radii at the center and at the top (and bottom) of the
BZ are in the same proportions as in the real FS, although in Fig. \ref{fig:FS}(b) the
distance between them has been enhanced for clarity.
The normalized conductance expected in the case of current injection
along the $ab$ plane can be calculated by using eq. 9 of Ref.
\cite{daghero11}. Note that, due to the particular mathematical form
of the equation, the relative position of the two FS sheets is
irrelevant. Of course, the choice of the symmetry of the order
parameters that reside on the two hyperboloids is determined by the
experiment. In this case, the absence of zero-bias maxima (in
\emph{all} the spectra we measured) suggests that both the gaps are
likely to be nodeless, in agreement with ARPES measurements \cite{terashima09}.
We therefore used in our model two $s$-wave
gaps whose amplitudes are pictorially indicated as gridded surfaces
in Fig. \ref{fig:FS}(b). We will call $\Delta_1$ and $\Delta_2$ the
large and the small gap, respectively. The 3D BTK model contains 3
adjustable parameters for each band, i.e. the gap amplitude
$\Delta$, the broadening parameter $\Gamma$ and the barrier
parameter $Z$. The weight of each band in the conductance is not
adjustable (as instead happens in the 2D model) being automatically
determined by the geometry of the FS and by the barrier parameter
$Z$. This considerably reduces the uncertainty on the best-fitting
parameters.

Fig. \ref{fig:Bafit} shows two examples of fit (lines) to the
experimental spectra. The values of the parameters are indicated in
the caption; the average values are $\Delta_1=11.75 \pm 0.25$ meV
and $\Delta_2=3.8 \pm 0.8$. In the 2D-BTK fit of the same curves the
gaps turned out to be $\Delta_1=10.7 \pm 0.2$ meV and $\Delta_2=4.4
\pm 0.6$ meV. With respect to the 2D BTK fit, the 3D one gives
smaller values of the small gap and larger values of the large gap,
and the theoretical curves are narrower. This is due to the fact
that in the 3D model the weight is fixed (in these two cases, the
weight of band 1 is about 0.2). Indeed, the 2D fit can be forced to
follow the experimental curve at energies higher than 13 meV (where
the 3D fit fails) if the weight of the bands is kept around 0.5, but
this clearly would not reflect the real shape of the FS sheets. The
inability of the model to reproduce the higher-energy structures (in
particular the kinks at about 20 meV) simply confirms that these
structures are related to effects that are not accounted for by the
model (and indeed can be explained as being due to the
strong electron-boson coupling, as shown in Refs.
\cite{tortello10,daghero11}).

\subsection{Ca(Fe$_{1-x}$Co$_x$)$_2$As$_2$ single crystals}
\begin{figure}[b]
\includegraphics[width=0.8\columnwidth]{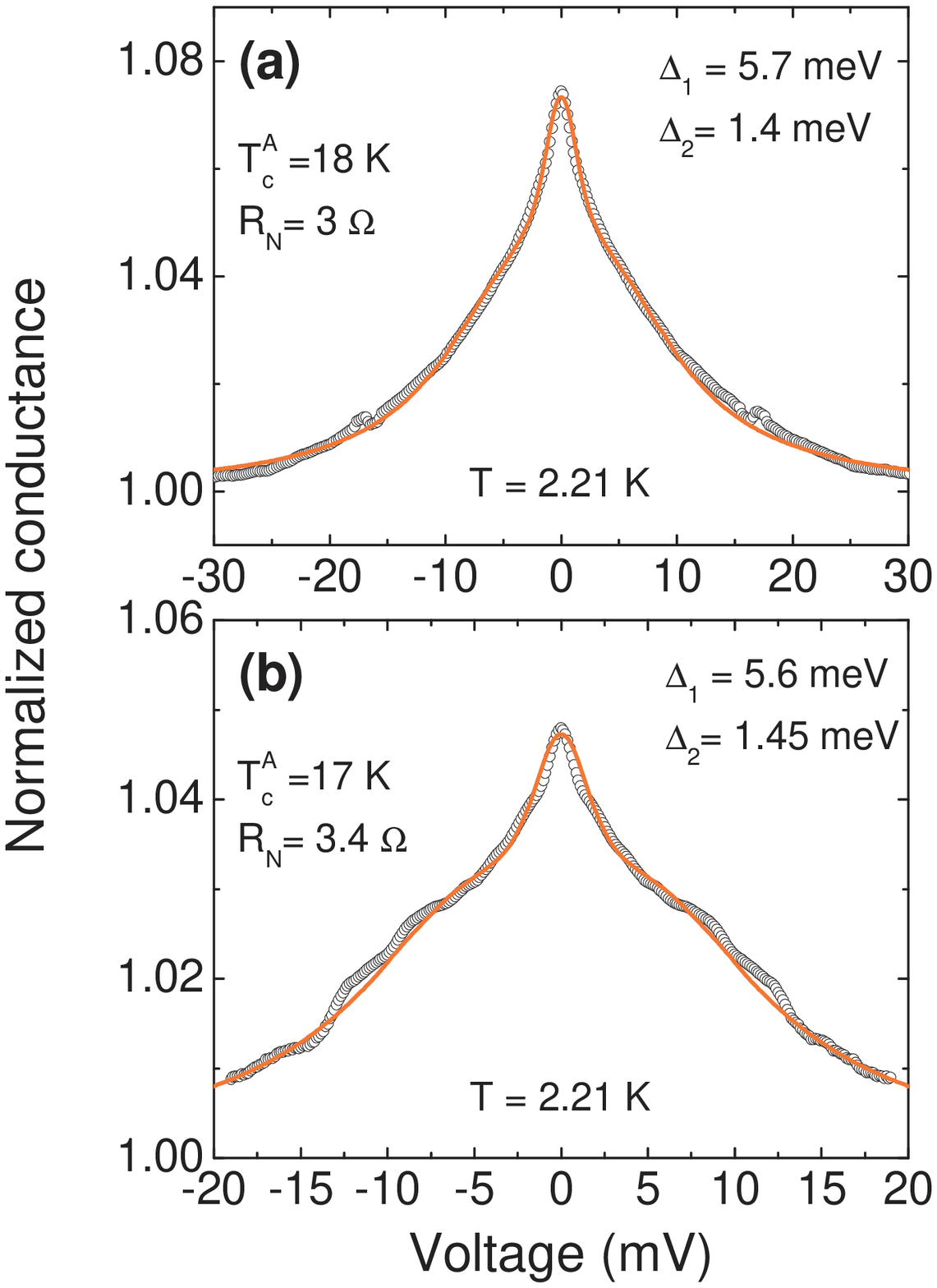}
\caption{Two examples of normalized conductance curves of $ab$-plane
point contacts in $\mathrm{Ca(Fe_{1-x}Co_x)_2As_2}$ single crystals
with $x=0.06$ (symbols) and of the relevant fit (lines) with the 3D BTK
model using the FS shown in Fig. \ref{fig:FS}(d). The fitting
parameters of the curve in (a) are: $\Delta_1=5.7$ meV,
$\Gamma_1=6.15$ meV, $Z_1=0.145$ and $\Delta_2=1.4$ meV,
$\Gamma_2=1.25$ meV, $Z_2=0.05$. The same parameters for the fit of
the curve in (b) are instead $\Delta_1=5.6$ meV, $\Gamma_1=6.5$ meV,
$Z_1=0.23$ and $\Delta_2=1.45$ meV, $\Gamma_2=1.7$ meV, $Z_2=0.05$.
In both (a) and (b) the angle between the normal to the interface
and the $a$ axis is $\alpha=\pi/8$.}\label{fig:Cafit}
\end{figure}

Figure \ref{fig:Cafit} shows two examples of typical conductance
curves of $ab$-plane contacts in $\mathrm{Ca(Fe_{1-x}Co_x)_2As_2}$
single crystals. Contrary to what happens in Co-doped Ba-122, here 100\% of
the conductance curves presents zero-bias maxima or peaks. As shown
elsewhere \cite{daghero11}, this is a clear sign that one of the
gaps is strongly anisotropic in the $(k_x k_y)$ plane. According to
theoretical calculations \cite{suzuki11}, nodal lines can appear in
the order parameter of the outer holelike FS sheet when the size of
the latter is largely increased in the vicinity of the top and
bottom faces of the BZ.  In Ref. \cite{suzuki11} this effect is
produced by the reduction in the pnictogen height $h_{As}$ induced
by the substitution of As with P. Our DFT calculations show that the
effect of Co substitution in $\mathrm{CaFe_2As_2}$ is very similar,
and that the FS of Fig. \ref{fig:FS}(c) is actually the extreme
consequence of a doping-induced increase in the warping of the
holelike sheet. Thus, our PCAR measurements confirm that, even
within a general $s\pm$ picture of spin-fluctuation mediated
superconductivity, nodal lines can appear in the holelike FS when
the latter is strongly deformed -- in particular, if it undergoes a
topological transition and splits into separate closed pockets.

We thus modeled the FS with one hyperboloid (for the electronlike
sheets) and one spheroid (for the holelike pocket) as in Fig.
\ref{fig:FS}(d) and assumed an isotropic large gap $\Delta_1$ on the
former and an anisotropic small gap $\Delta_2$ on the latter. This
gap should represent an evolution of that depicted in Refs.
\cite{suzuki11} and \cite{graser10} when the relevant FS sheet
becomes a closed surface. One possibility is to use a $d$-wave gap
$\Delta_2(\theta, \phi)=\Delta_2 \cos(2\theta)\sin(\phi)$ as in
Ref. \cite{gonnelli12}, to reproduce the sign change of the OP.
However, the symmetry of the OP of Refs. \cite{suzuki11} and
\cite{graser10} in the $(k_x k_y)$ plane is different from the
$d$-wave one and features two additional axes of equation $k_y=\pm
k_x$ like the $s+g_{x^2-y^2}$ one \cite{vanharlingen}. As a
consequence, the probability of constructive interference between
electronlike and holelike quasiparticles (that gives rise to the
zero-bias peak) is strongly reduced.  However, the zero-bias maximum
could also arise from the existence of angular regions where the gap has
very small amplitude, rather than from its change of sign, as shown
in Ref. \cite{daghero11}. To account for this possibility, here we
modeled the gap on the holelike pockets with a fully anisotropic
$s$-wave gap of equation $\Delta_2(\theta, \phi)=\Delta_2 \cos^4
(2\theta)\sin(\phi)$ (gridded surface in Fig. \ref{fig:FS}(d)). The
fit of the experimental curves is shown in Fig. \ref{fig:Cafit}
(solid lines). The parameters of these two particular fits are
listed in the caption. From different fits we get the following
average values for the gaps: $\Delta_2=1.4 \pm 0.1$ meV and
$\Delta_1=5.5 \pm 0.3$ meV. If one chooses a $d$-wave symmetry for
the small gap, a fit of comparable quality is obtained but the
values of the gaps are $\Delta_1=1.6 \pm 0.1$ meV and $\Delta_2= 5.3
\pm 0.2$ meV \cite{gonnelli12}.

\subsection{Ba(Fe$_{1-x}$Co$_x$)$_2$As$_2$ films}
\begin{figure}[b]
\includegraphics[width=0.8\columnwidth]{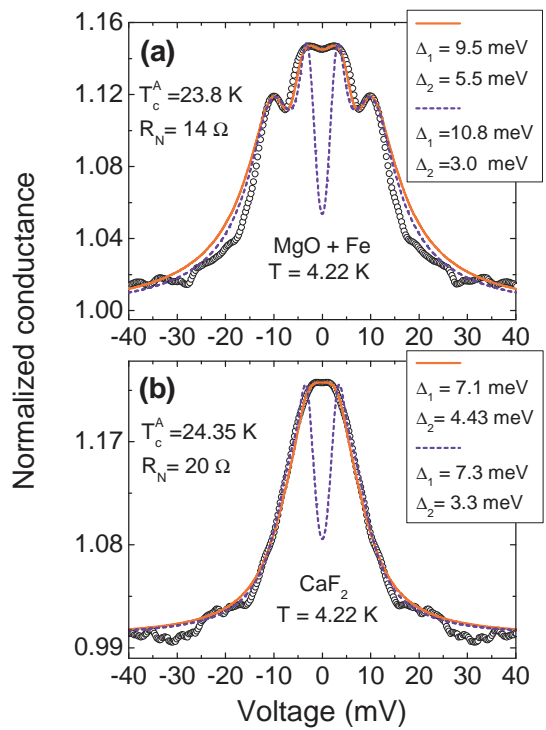}
\caption{Two examples of normalized conductance curves of $c$-axis
point contacts in $\mathrm{Ba(Fe_{1-x}Co_x)_2As_2}$ films with
$x=0.08$ (symbols) but on different substrates: MgO + Fe buffer
layer in (a) and CaF$_2$ with no buffer layer in (b). The dashed
lines represent the 3D BTK fit, while solid lines represent the 2D
BTK fit. The corresponding gap amplitudes are indicated in the
labels. The weights of the bands in the 3D fit are $w_1=0.39$ and
$w_2=0.61$. The 2D fit was performed by fixing instead $w_1=0.44$
and $w_2=0.56$.}\label{fig:Bafilm}
\end{figure}

Two examples of PCAR spectra in $\mathrm{Ba(Fe_{1-x}Co_x)_2As_2}$
films are shown in Fig. \ref{fig:Bafilm}. The spectrum in (a) refers
to a film on MgO substrate with Fe buffer layer, and shows
double-gap features similar to those of Fig. \ref{fig:Bafit}(b) but
even clearer. The spectrum of panel (b) was measured in a film on
CaF$_2$ and presents much smoother structures. In particular, the
large gap manifests itself in very broad shoulders at about $\pm 6$ meV.
As stated in a previous section, the normalization in this case gives rise to
some ambiguity since the normal state at $T_c^A$ is not related to
the contact alone, but includes a contribution from the portion of
the film between the point contact and the voltage electrode. The
normalization by a guess normal state can be used, but one should
keep in mind that the choice of the presumed normal state can affect
the shape of the curve and, to some extent, the fitting parameters
-- actually, the broadening $\Gamma$ and the barrier parameter $Z$
are the most affected by that choice while the values of the gaps
are rather robust. The dashed lines in Fig. \ref{fig:Bafilm}
represent the results of the 3D BTK fit using the same model FS as
in Fig. \ref{fig:FS}(a).
The 3D model is clearly unable to fit the curve very well because of
a $Z$-enhancing effect due to the shape of the FS \cite{daghero11}.
The parameters of the fit shown in Fig. \ref{fig:Bafilm}(a) are
$\Delta_1=10.8$ meV, $\Gamma_1=3.85$ meV, $Z_1=0.1$, $\Delta_2=3.0$
meV, $\Gamma_2=1.4$ meV, $Z_2=0$. The weights of the bands, which
are not adjustable within this model, are $w_1=0.39$ and $w_2=0.61$.
Similarly, in Fig. \ref{fig:Bafilm}(b) the 3D BTK fit (dashed curve) was
obtained by using $\Delta_1=7.3$ meV, $\Gamma_1=2.65$ meV, $Z_1=0.02$,
$\Delta_2=3.3$ meV, $\Gamma_2=1.2$ meV, $Z_2=0$. Note that in both
cases a very deep zero-bias minimum is obtained despite the fact
that the intrinsic barrier strength is put equal to zero in band $2$
and is very small in band $1$. The failure of the 3D BTK model may
be due to different factors. One possibility is that, for some
reasons, the probe current is injected mainly along the $ab$ plane
despite the orientation of the film. This may happen, for example,
when the surface is rough, a minority percentage of grains are not
oriented or the point contact is made by using a sharp tip that may
pierce the surface. In our case, all these possibilities are rather
unlikely. The other possibility is that the use of two isotropic
gaps is not the best choice to reproduce the real distribution of
the gaps over the FS. Indeed, the presence of ``hot spots'' where
the gap is significantly suppressed has been evidenced by Raman
spectroscopy \cite{muschler09} and then justified theoretically
\cite{mazin10}. Since the hot spots seem to reside on the
electronlike FS sheets, one might try to improve the model by using
a small anisotropic gap with zeros and a large isotropic gap (as we did in the case of
$\mathrm{Ca(Fe_{1-x}Co_x)_2As_2}$). This would certainly improve the fit
because there would be quasiparticle excitations even at very low
energy and the Z-enhancing effect might be compensated by the
zero-bias maximum due to these excitations \cite{daghero11}. We will
explore this possibility elsewhere. For the time being, we show in
Fig. \ref{fig:Bafilm} what would be the result of a 2D-BTK fit
(solid lines). The fit is rather good in the energy region of the
gaps and catches the main features of the curves, but just because
the weights of the bands were taken to be different from those
determined by the shape of the FS in the 3D fit. In (a) we got
$\Delta_1=9.5$ meV and $\Delta_2=5.5$ meV by using $w_1=0.44$ and
$w_2=0.56$. Of course, changing the weight affects all the fitting
parameters: the resulting gap values (averaged over different fits)
are $\Delta_1=9.5 \pm 0.2$ meV and $\Delta_2=5.4 \pm 0.2$ meV. The
uncertainty is small because the peaks are rather sharp. In (b) we
obtained $\Delta_1=7.1$ meV and $\Delta_2=4.43$ meV, with the
weights $w_1=w_2=0.5$. Averaging over different fits obtained by
changing the weight, the gap values turn out to be $\Delta_1=7.4 \pm
0.3$ meV and $\Delta_2=4.2 \pm 0.25$. Note that the 2D-BTK fit of
similar curves (with current injection along the $c$ axis) in single
crystals of the same material gave $\Delta_1=9.2 \pm 1.0$ meV and
$\Delta_2=4.1 \pm 0.4$ meV \cite{tortello10}. There is thus a
partial superposition of the gap values, although not a perfect
agreement. This is not particularly surprising since: i) there may
be some effect of the substrate, especially in the case of the films
on Fe buffer layer; ii) the measurements in films highlight a
certain degree of inhomogeneity in the superconducting properties,
although the film itself is of very high quality; iii) the
statistics of the measurements in films needs to be extended to draw
definite conclusions.

\section{Conclusions}
We have presented the results of PCAR experiments in
$\mathrm{Ba(Fe_{1-x}Co_x)_2As_2}$ single crystals and films with
$x=0.08$ and in $\mathrm{Ca(Fe_{1-x}Co_x)_2As_2}$ single crystals
with $x=0.06$. The results of these measurements have been analyzed
by using a novel 3D version of the BTK model that accounts for the
shape of the FS -- actually using an analytical model for the real FS
calculated within DFT. The results indicate that the two systems,
although belonging to the same 122 family of Fe-based compounds,
show major differences in the shape of the spectra, and this is a
symptom of a difference in the symmetry of the order parameters.  In
$\mathrm{Ca(Fe_{1-x}Co_x)_2As_2}$ a zero-bias peak or maximum is
always observed and can be explained as being due to a strong
anisotropy of the smaller gap. DFT calculations show that, at 8\% Co
content, the holelike FS sheet undergoes a topological transition
and splits into separated closed pockets. This can be interpreted as
an extreme consequence of the increasing-with-doping warping of this
FS sheet, which is predicted to be accompanied by the emergence of
3D nodes in the relevant gap. Indeed, a fit of the spectra is
possible by assuming either a $d$-wave \cite{gonnelli12} or an
anisotropic symmetry of the small gap, to mimic the possible
evolution of these 3D nodes when the topological transition occurs.
In the case of $\mathrm{Ba(Fe_{1-x}Co_x)_2As_2}$ at $x=0.08$, the
calculated FS shows markedly warped holelike sheets but no
topological transitions. The spectra taken in single crystals with
the current injected along the $ab$ plane can be fitted with two
isotropic gaps; the agreement between model and data is limited to
the central region of the spectra because of additional structures
around 20 meV that are due to the strong electron-boson coupling
\cite{tortello10,daghero11} and cannot be reproduced within this
approach. Instead, the fit of the $c$-axis spectra measured in thin
films with the same model is unsuccessful around zero bias, where
the so-called $Z$-enhancing effect due to the shape of the FS
produces a strong depression of the conductance which is not
observed experimentally. This might be the sign of the presence of
gap minima \cite{mazin10} whose existence has been inferred  from
Raman spectroscopy results \cite{muschler09}.

\section*{Acknowledgments}
This work was done under the Collaborative EU-Japan Project ``IRON
SEA'' (NMP3-SL-2011-283141) and under the PRIN Project No.
2008XWLWF9-005 of the Italian Ministry of Research.


\begin{thebibliography}{10}

\bibitem{kamihara08}
Y.~Kamihara, T.~Watanabe, M.~Hirano, and H.~Hosono.
\newblock {\em J. Am. Chem. Soc.}, 130:3296, 2008.

\bibitem{mazin08}
I.~I. Mazin, D.~J. Singh, M.~D. Johannes, and M.~H. Du.
\newblock {\em Phys. Rev. Lett.}, 101:057003, 2008.

\bibitem{mazin10b}
I.~I. Mazin.
\newblock {\em Nature}, 464:183, 2010.

\bibitem{paglione10}
J.~Paglione and R.~L. Greene.
\newblock {\em Nature Phys.}, 6:645--58, 2010.

\bibitem{mazin10}
I.~I. Mazin, T.~P. Devereaux, R.~Hackl, B.~Muschler, J.~G. Analytis,
Jiun-Haw
  Chu, and I.~R. Fisher.
\newblock {\em Phys. Rev. B}, 82:180502(R), 2010.

\bibitem{kuroki09}
K.~Kuroki, H.~Usui, S.~Onari, R.~Arita, and H.~Aoki.
\newblock {\em Phys. Rev. B}, 79:224511, 2009.

\bibitem{suzuki11}
K.~Suzuki, H.~Usui, and K.~Kuroki.
\newblock {\em J. Phys. Soc. Jpn.}, 80:013710, 2011.

\bibitem{terashima09}
K.~Terashima, J.~H. Bowen, K.~Nakayama, T.~Sato, P.~Richard, Y.-M.
Xu, L.~J.
  Li, G.~H. Cao, Z.-A. Xu, H.~Ding, and T.~Takahashi.
\newblock {\em Proc. Natl. Acad. Sci (USA)}, 106:7330, 2009.

\bibitem{gonnelli12}
R.~S. Gonnelli, M.~Tortello, D.~Daghero, R.~K. Kremer, Z.~Bukovski,
N.~D.
  Zhigadlo, and J.~Karpinski.
\newblock {\em Supercond. Sci. Technol.}, 25:065007, 2012.

\bibitem{muschler09}
B.~Muschler, W.~Prestel, R.~Hackl, T.~P. Devereaux, J.~G. Analytis,
J.-H. Chu,
  and I.~R. Fisher.
\newblock {\em Phys. Rev. B}, 80:180510(R), 2009.

\bibitem{sefat08b}
A.~S. Sefat, R.~Jin, M.~A. McGuire, B.~C. Sales, D.~J. Singh, and
D.~Mandrus.
\newblock {\em Phys. Rev. Lett.}, 101:117004, 2008.

\bibitem{matusiak10}
M.~Matusiak, Z.~Bukowski, and J.~Karpinski.
\newblock {\em Phys. Rev. B}, 81:020510(R), 2010.

\bibitem{harnagea11}
L.~Harnagea, S.~Singh, G.~Friemel, N.~Leps, D.~Bombor,
M.~Abdel-Hafiez,
  A.~U.~B. Wolter, C.~Hess, R.~Klingeler, G.~Behr, S.~Wurmehl, and B.~B¨uchner.
\newblock {\em Phys. Rev. B}, 83:094523, 2011.

\bibitem{daghero10}
D. Daghero and R.S. Gonnelli.
\newblock {\em Supercond. Sci. Technol}, 23:043001, 2010.

\bibitem{chen10b}
T.~Y. Chen, S.~X. Huang, and C.~L. Chien.
\newblock {\em Phys. Rev. B}, 81:214444, 2010.

\bibitem{gonnelli09a}
R.~S. Gonnelli, D.~Daghero, M.~Tortello, G.~A. Ummarino, V.~A.
Stepanov, J.~S.
  Kim, and R.~K. Kremer.
\newblock {\em Phys. Rev. B}, 79:184526, 2009.

\bibitem{singh08b}
D.~J. Singh.
\newblock {\em Phys. Rev. B}, 78:094511, 2008.

\bibitem{hu11}
R.~Hu, S.~Ran, S.L. Bud'ko, W.~E. Straszheim, and P.~C. Canfield.
\newblock Unpublished, arXiv:1111.7034, Nov 2011.

\bibitem{colonna11}
N.~Colonna, G.~Profeta, A.~Continenza, and S.~Massidda.
\newblock {\em Phys. Rev. B}, 83:094529, 2011.

\bibitem{mittal11}
R.~Mittal, S.~K. Mishra, S.~L. Chaplot, S.~V. Ovsyannikov,
E.~Greenberg, D.~M.
  Trots, L.~Dubrovinsky, Y.~Su, Th. Brueckel, S.~Matsuishi, H.~Hosono, and
  G.~Garbarino.
\newblock {\em Phys. Rev. B}, 83:054503, 2011.

\bibitem{tortello10}
M.~Tortello, D.~Daghero, G.~A. Ummarino, V.~A. Stepanov, J.~Jiang,
J.~D. Weiss,
  E.~E. Hellstrom, and R.~S. Gonnelli.
\newblock {\em Phys. Rev. Lett.}, 105:237002, 2010.

\bibitem{daghero11} D.~Daghero, M.~Tortello, G.A. Ummarino, and
R.~S. Gonnelli.
\newblock {\em Rep. Prog. Phys.}, 74:124509, 2011.

\bibitem{graser10}
S.~Graser, A.~F. Kemper, T.~A. Maier, H.-P. Cheng, P.~J. Hirschfeld,
and D.~J.
  Scalapino.
\newblock {\em Phys. Rev. B}, 81:21450, 2010.

\bibitem{vanharlingen}
D.~J. Van~Harlingen.
\newblock {\em Rev. Mod. Phys}, 67:515--37, 1995.

\end{thebibliography}

\end{document}